\documentclass[twocolumn,           
               showpacs,            
               nopreprintnumbers,   
               aps,                 
               prd,          	    
               a4paper,             
               groupedaddress,      
               unsortedaddress,
               nofootinbib,         
               tightenlines,        
               floats               
               ]{revtex4}

\usepackage{graphicx}
\usepackage{amssymb,latexsym}
\usepackage[draft=false]{hyperref}

\begin{document}




\title{Nflation: multi-field inflationary dynamics and perturbations}
\author{Soo A Kim and Andrew R.~Liddle}
\affiliation{Astronomy Centre, University of Sussex, 
             Brighton BN1 9QH, United Kingdom}
\date{\today} 
\pacs{98.80.Cq \hfill astro-ph/0605604}
\preprint{astro-ph/0605604}


\begin{abstract}
We carry out numerical investigations of the dynamics and
perturbations in the Nflation model of Dimopoulos et al.~(2005). This
model features large numbers of scalar fields with different masses,
which can cooperate to drive inflation according to the assisted
inflation mechanism. We extend previous work to include random initial
conditions for the scalar fields, and explore the predictions for
density perturbations and the tensor-to-scalar ratio. The
tensor-to-scalar ratio depends only on the number of $e$-foldings and
is independent of the number of fields, their masses, and their
initial conditions.  It therefore always has the same value as for a
single massive field.  By contrast, the scalar spectral index has
significant dependence on model parameters. While normally multi-field
inflation models make predictions for observable quantities which
depend also on the unknown field initial conditions, we find evidence
of a `thermodynamic' regime whereby the predicted spectral index
becomes independent of initial conditions if there are enough
fields. Only in parts of parameter space where the mass spectrum of
the fields is extremely densely packed is the model capable of
satisfying the tight observational constraints from WMAP3
observations.
\end{abstract}

\maketitle

\section{Introduction}

Dimopoulos et al.~\cite{DKMW} have recently described an interesting
addition to the collection of known inflationary models with
motivation from particle theory. They consider the many axion fields
predicted by string vacuum solutions, and show that these fields can
work cooperatively to drive a period of inflation, via the assisted
inflation mechanism \cite{LMS}, without any of those fields needing to
take on values in excess of the Planck scale. This ensures that the
theory can be considered in the regime of radiative stability.

The prediction is that there will be large numbers of fields, all with
different masses. Such a setup was first considered by Kanti and Olive
\cite{KO} and further explored by Kaloper and Liddle \cite{KL}, in the
context of Kaluza--Klein models where there would be a tower of mass
eigenstates. In the axion proposal there is also a spectrum of masses
states, the main difference being that these may be very closely
packed. Related ideas are discussed in Ref.~\cite{mfield}.

Dimopoulos et al. made only a rudimentary study of the dynamics,
assuming that all fields would begin with the Planck value and that
all fields would slow-roll together. This leads to a prediction for
density perturbations that matches that of a single massive
field. Subsequently, an interesting generalization was made by Easther
and McAllister \cite{EM}, who used results from random matrix theory
to predict the likely distribution of field masses. Their main
investigation was still restricted, however, to quite specific choices
of initial conditions for the fields.

Our main aim in this paper is to consider more realistic initial
conditions, where each field starts in a random location within the
sub-Planckian regime. We will show that this typically reduces the
amount of inflation achieved, necessitating a greater number of
fields. The most interesting consequence, however, concerns the
perturbations that arise. Ordinarily multi-field inflation predictions
depend on the initial conditions chosen (something we will verify
explicitly in the two-field case) and so the models cannot be said to
make definite predictions for observations, unlike the single-field
case. However, focussing on the adiabatic perturbations, we find once
the fields become sufficiently densely packed, the predictions once
more become independent of the field initial conditions. This is
essentially analogous to gas thermodynamics; once there are enough
fields they probe the space of possible initial conditions well enough
that the ensemble of fields can be described in a collective
fashion. Even more strikingly, we find a completely model-independent
prediction for the tensor-to-scalar ratio, confirming a result of
Alabidi and Lyth \cite{AL}.

\section{Multi-field dynamics and Nflation}

Phenomenologically, the basic set-up of Nflation is straightforward
--- a set of massive uncoupled scalar fields with a particular mass
spectrum. While any one of those fields is able separately to drive
inflation provided it is displaced sufficiently from its minimum, if
one imposes the restriction that the field value does not exceed the
reduced Planck mass $M_{{\rm Pl}}$ then single-field inflation is no
longer possible in such potentials. Dimopoulos et al.~\cite{DKMW}
noted that this situation can be saved provided there are enough
fields, through the assisted inflation phenomenon \cite{LMS}.

Assisted inflation is simply the realization that in multi-field
systems, each field feels the downward force from its own potential,
but the collective frictional force from all the fields, and hence
slow-roll is more easily achieved. In the original work of Liddle et
al.~\cite{LMS} exponential potentials were studied, in which the
assisted behaviour was reinforced by the presence of a late-time
attractor where all fields contribute to the evolution. In the case of
massive fields there is no such attractor solution, but the generic
phenomenon of enhanced friction remains. With enough fields, it
becomes possible to drive sufficient inflation without any field
exceeding the reduced Planck mass.

For a set of uncoupled fields, the equation for the number of
$e$-foldings is
\begin{eqnarray}
N & \simeq &  -\frac{1}{M_{{\rm Pl}}^2} \sum_j
\int_{\phi_j}^{\phi_j^{{\rm end}}} \frac{V_j}{V'_j} \, d\phi_j \,;\\
& \simeq & \frac{\sum_j \phi_j^2}{4M_{{\rm Pl}}^2} \,.
\label{e:efolds1} 
\end{eqnarray}
Here $V_j$ is the potential of the $j$-th field $\phi_j$,
$V'_j \equiv dV_j/d\phi_j$, and throughout there are no
summations unless indicated explicitly.  The last line specializes to
the Nflation case of massive uncoupled fields. In this case the lower
limits of the integrals, corresponding to the end of inflation, can be
neglected and have been.

\begin{figure}[t]
\centering
\includegraphics[width=7.5cm]{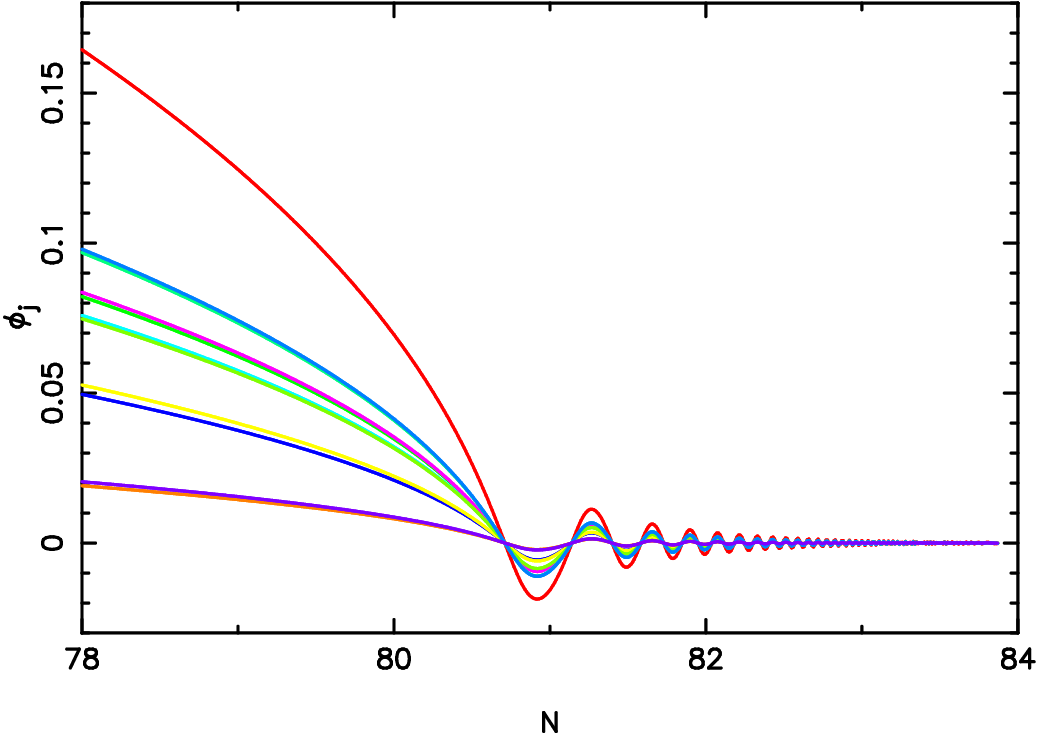}
\caption{\label{f:me} The evolution of eleven of the fields in a
1000-field simulation, where the field initial conditions are randomly
chosen. The simulation started at $N=0$.}
\end{figure}

\subsection{The equal mass case}

In the very simplest case where all fields have the same mass, there
is a different type of attractor corresponding to radial motion in
field space. Dimopoulos et al. assumed the fields all started with the
same initial condition, hence automatically placing them on this
radial trajectory. With initial condition $\phi_j = \alpha M_{{\rm
Pl}}$ they found the total number of $e$-foldings was $N_{{\rm total}}
= \alpha^2 N_{{\rm f}}/4$ where $N_{{\rm f}}$ is the total number of
fields, as indeed follows immediately from Eq.~(\ref{e:efolds1}).

We instead use random initial conditions, with the field initial
conditions chosen uniformly in the range $(0, M_{{\rm Pl}})$ [which is
equivalent by symmetry to $(-M_{{\rm Pl}},M_{{\rm Pl}})$]. When the
fields all have the same mass they quickly adopt a radial trajectory;
they all then exit the slow-roll regime simultaneously and oscillate
in phase --- see Fig.~\ref{f:me}.  We find that the total number of
$e$-foldings achieved can be well approximated by
\begin{equation}
\label{e:efolds}
N_{{\rm total}} \simeq \frac{N_{{\rm f}}}{12} \,.
\end{equation}
This indeed follows from Eq.~(\ref{e:efolds1}) since for our initial
conditions $\langle \phi_j^2 \rangle = M_{{\rm Pl}}^2/3$.  This
indicates that they overestimated the number of $e$-foldings achieved
by a factor of about three through not using random initial
conditions.  At least 600 fields are required to give the minimum
acceptable amount of inflation if none of them exceeds the reduced
Planck mass.

Dimopoulos et al. were also able to show that the adiabatic
perturbations on this radial trajectory have the same spectral index
as in the single-field case.  This was subsequently confirmed in a
more detailed analysis by Byrnes and Wands \cite{BW}, who also noted
the additional possibility of a large number of scale-invariant
isocurvature perturbations from the tangential directions, which might
or might not become important depending on the subsequent
evolution. 

\subsection{A mass spectrum}

For definiteness, throughout we will consider the mass spectrum
suggest by Dimopoulos et al, where the fields are distributed
exponentially in mass. We write the hierarchy slightly differently,
normalizing to the lightest mass $m$, writing
\begin{equation}
m_j^2 = m^2 \exp (j/\sigma) \quad j=0,\cdots ,N_{{\rm f}}-1 \,.
\end{equation}
Here $\sigma$ gives the density of fields per logarithmic mass
interval. If one were to decide that the heaviest field should have
the reduced Planck mass, as in Ref.~\cite{DKMW}, then that would
impose a relation amongst $m$, $\sigma$ and $N_{{\rm f}}$.

Easther and McAllister \cite{EM} have argued that random matrix theory
predicts a somewhat different form for the mass spectrum, known as the
Mar\u{c}enko--Pastur law. Investigation of this more complicated form is
beyond the scope of our present paper, and we do not expect it to lead
to significant qualitative differences, but a detailed investigation
would nevertheless be interesting.

\begin{figure}[t]
\centering
\includegraphics[width=7.5cm]{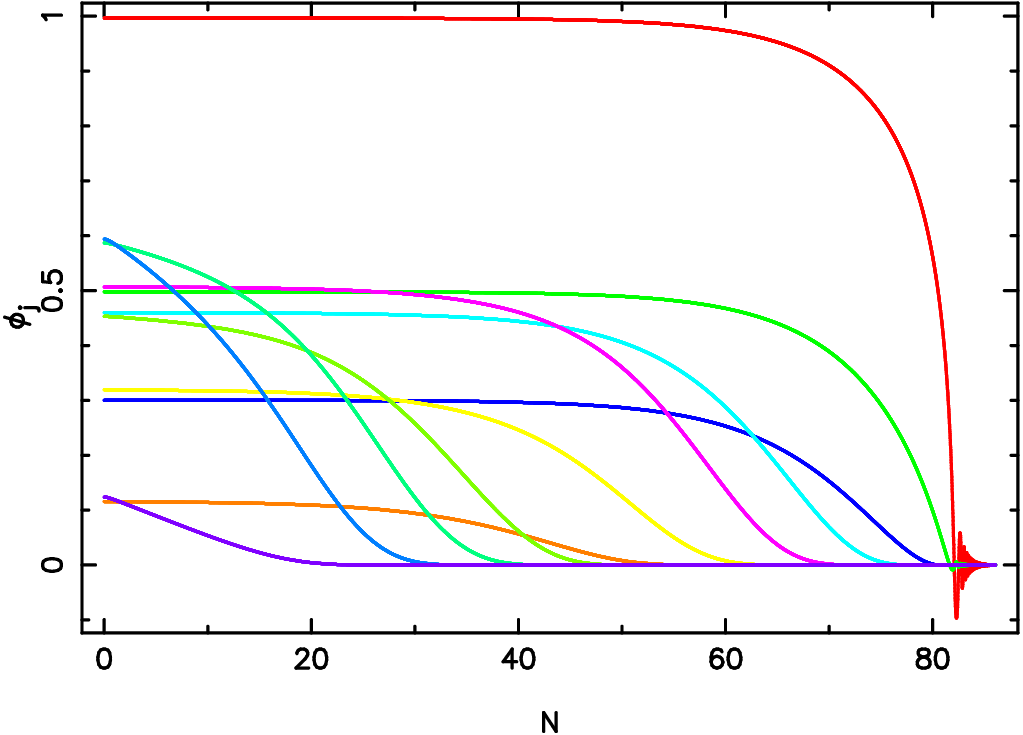}
\caption{\label{f:multi} The evolution of eleven fields from a 1000-field
simulation with different masses, with $\sigma = 100$ and $N_{{\rm f}}
= 1000$. The fields shown are $j=0,100,\cdots,900$ and $999$. The more
massive fields exit slow-roll first.}
\end{figure}

For collections of fields with different masses, within the slow-roll
regime the fields obey the condition \cite{KO}
\begin{equation}
\frac{\phi_j}{\phi_{j,0}} = \left(\frac{\phi_i}{\phi_{i,0}}
\right)^{m_j^2/m_i^2} \,.
\end{equation}
where $i$ is some particular index, $j$ runs over all the fields, and
the subscript `0' indicates initial value.  This indicates that they
retain a memory of their initial conditions --- there is no global
attractor. Further, different fields will exit slow-roll at different
times; typically the high-mass fields will reach their minima first,
except by chance effects of initial conditions. This has the important
consequence that the dynamics of the final stages of inflation is
usually determined by some number of the lightest fields in the
ensemble, with the effects of any heavier fields vanishing before our
observable Universe exits the horizon. Figure~\ref{f:multi} shows a
particular example.

Dimopoulos et al.~did not make any detailed exploration of the
dynamics in the unequal mass case, but some investigation was made by
Easther and McAllister \cite{EM}. They considered two choices of
initial condition, by which they mean the configuration at 60
$e$-foldings before the end of inflation. The first was to take the fields
to have initially the same field value, and the second to take them as
having the same energy density. In either case, this `initial' 
property is swiftly destroyed by the dynamical evolution.

We instead choose random initial conditions for the fields as
described above, and compute the total number of $e$-foldings as a
function of the parameters $\sigma$ and $N_{{\rm f}}$.  We find that
for $\sigma \gtrsim N_{{\rm f}}/10$, the amount of inflation becomes
independent of $\sigma$ and is well approximated by Eq.~\ref{e:efolds}
above. This confirms that of order 1000 fields are needed if all
sub-Planckian. For smaller $\sigma/N_{{\rm f}}$ the fields become
spread across several orders of magnitude in mass and numerical
simulation becomes difficult, the expectation however being that the
more massive fields fall rapidly to their minimum and play no further
role before the lighter fields move significantly.

\section{Perturbations}

We evaluate the perturbation spectrum using the formalism of Sasaki
and Stewart \cite{SS}, who showed that the perturbation spectrum of
the curvature perturbation ${\cal R}$ at the end of inflation is given
by
\begin{equation}
{\cal P}_{\cal R} = \left(\frac{H}{2\pi} \right)^2 \,
\frac{dN}{d\phi_i} \, \frac{dN}{d\phi_j} \, \delta_{ij} \,,
\end{equation}
where we follow the notation of Ref.~\cite{LL}, $N$ again being the
number of $e$-foldings. 

In general, evaluation of this expression is quite tricky, as one
should compute the change in the number of $e$-foldings by tracing the
trajectory past the end of inflation and through reheating until a
fixed density during the radiation era. However provided the
multi-field inflation adopts a straight trajectory in field space
before inflation ends, the curvature perturbation will already become
constant at that time. This would happen in Nflation provided the last
few $e$-foldings are driven by only one field, which is likely but not
inevitable.

We will approximate the formula using the slow-roll approximation,
according to which it can be 
written \cite{LR,EM} 
\begin{eqnarray}
{\cal P}_{\cal R} & \simeq & \frac{V}{12\pi^2 M_{{\rm Pl}}^6}  \sum_j
\left( \frac{V_j}{V'_j} \right)^2 \,; \\
 & \simeq & \frac{\sum_k m_k^2 \phi_k^2}{96\pi^2 M_{{\rm Pl}}^6}
\sum_j \phi_j^2 \,,
\end{eqnarray}
where $V \equiv \sum_k V_k$ is the total potential, and the second
line specializes to the Nflation case. An interesting observation is
that for Nflation the power spectrum is proportional to $H^2 N$,
regardless of initial conditions.  In principle the summation should
be only over those fields which are slowly-rolling, though we find in
practice that the contribution from non-slow-rolling fields is
negligible. The above expressions depend only on quantities at
horizon crossing but nevertheless should offer a good approximation,
as we discuss in detail below for the two-field case.

The spectral index is then given by \cite{SS}
\begin{eqnarray}
\nonumber 
n_{{\rm S}}-1 &\simeq& -\frac{M_{\rm Pl}^2}{V^2}\sum_k (V'_k)^2 
-\frac{ 2M_{\rm Pl}^2 }{\sum_j\left( V_j/V'_j\right)^2} \\
&& + \frac{2M_{\rm Pl}^2}{V}\left[ \frac{\sum_l
\left({V_l}/{V'_l}\right)^2 V''_l}  
{\sum_j \left( {V_j}/{V'_j}\right)^2}\right]\,; \\
\label{e:index}
&\simeq& - 4 M_{\rm Pl}^2 \left[ \frac{\sum_k m_k^4 \phi_k^2}
{(\sum_l m_l^2 \phi_l^2)^2} + \frac{1}{\sum_j \phi_j^2}\right]\,,
\end{eqnarray}
where again the last line specializes to Nflation.

The other main inflationary observable is the tensor-to-scalar ratio
$r$. The tensor spectrum is given by the usual formula, as it depends
only on the expansion rate history $H(a)$,
\begin{equation}
{\cal P}_g = \frac{2 H^2}{\pi^2 M^2_{\rm Pl}}\simeq \frac{2V}{3\pi^2
M_{\rm Pl}^4}\,.  
\end{equation}
Thus the tensor-to-scalar ratio is given by
\begin{eqnarray}
r \cong \frac{{\cal P}_g}{{\cal P}_{\cal R}}
&\simeq& \frac{8 M_{\rm Pl}^2}{\sum_j \left({V_j}/{V'_j}\right)^2}\,;\\
&\simeq& \frac{32 M^2_{\rm Pl}}{\sum_j \phi_j^2} \simeq \frac{8}{N}\,,
\end{eqnarray}
with the last line holding only for the Nflation case of massive
coupled fields and which uses Eq.~(\ref{e:efolds1}).

This last formula immediately gives a striking result first noted by
Alabidi and Lyth \cite{AL}: in Nflation the tensor-to-scalar ratio is
\emph{independent} (within the slow-roll approximation) of the number
of fields present, their masses, and their initial conditions. It is
given simply by the number of $e$-foldings at which the expression is
evaluated. This result is confirmed by our numerical code, with
variation only in the third significant figure due to slow-roll
corrections.  Nflation therefore gives a definitive prediction for the
tensor-to-scalar ratio, and moreover one that is readily accessible to
upcoming experiments.

We will throughout assume that the observable scales crossed outside
the horizon 50 $e$-foldings before the end of inflation. For
comparison with later results, the observables obtained for a single
massive field are given by
\begin{eqnarray}
n_{{\rm S}}-1 & = & -0.04 \quad (-0.0404)\\
r & = & 0.16 \quad \; \; \: (0.162)
\end{eqnarray}
where the first number uses the slow-roll approximation both for the
spectrum and for computing the 50 $e$-foldings point, and the number
in brackets shows how this is corrected if the 50 $e$-foldings point
is computed numerically (as is the case for the multi-field results we
will display). The observed normalization of the spectrum can be taken
as ${\cal P}_{\cal R}^{1/2} = 5 \times 10^{-5}$ \cite{LL}, leading to
a normalization of
\begin{equation}
\frac{m}{M_{{\rm Pl}}} = 7.8 \times 10^{-6} \,,
\end{equation}
in the single-field case.

\subsection{The two-field case}

As a test of our code, we first study the case of two massive
fields. This was previously studied in the slow-roll approximation by
Lyth and Riotto \cite{LR}, and more recently using full trajectory
integration by Vernizzi and Wands \cite{VW}. 

Starting with the Lyth--Riotto result, they used the
slow-roll approximation for both fields and found that the spectral
index was given by
\begin{eqnarray}
\nonumber
n_{\rm S}-1 &\simeq&  -\frac{1}{N}\left[ 1+ \frac{(1+f)(1+R^4 f)}
{(1+R^2 f)^2} \right]\,;\\
\label{e:index2}
&=&  -\frac{1}{N}\left[ 2 + \frac{(R^2 -1)^2 f}{(1+R^2f)^2}\right]\,,
\end{eqnarray}
where $f \equiv \phi_2^2 / \phi_1^2$ and $R \equiv m_2 / m_1$. This is
indeed the two-field version of Eq.~(\ref{e:index}), using
Eq.~(\ref{e:efolds1}).  The prediction depends explicitly on $f$,
which is the ratio of the values of the two fields at 50 $e$-foldings,
showing clearly that this model makes no unique prediction for the
spectral index.

This expression has various symmetries. For instance, the spectral
index is independent of $f$ in the equal-mass case and takes on the
single-field value.  It is of course invariant under simultaneous
interchange $R \rightarrow 1/R$ and $f \rightarrow 1/f$, corresponding
to swapping the labels on the fields. The general form of the
correction to the single-field result is however quite complex, with
it becoming large in some parts of the $R$--$f$ plane.

A more sophisticated treatment of the two-field model was recently
given by Vernizzi and Wands \cite{VW}, who provide a formalism able to
track the evolution of the spectral amplitude and index during
inflation up until its final value. Their expression for the spectral
index mostly features terms evaluated at horizon crossing, plus one
additional term denoted $Z_c$. This term accounts for the contribution
to the change in $e$-foldings at the final uniform-density
hypersurface, and evolves during inflation driving evolution of
$n_{{\rm S}}$. If $Z_c$ is set to zero, our formula
Eq.~(\ref{e:index}) is recovered. We have reproduced their
calculation, and find that while $Z_c$ is substantial at horizon
crossing, it becomes negligible by the end of inflation. Accordingly,
our expression is an excellent approximation to the desired answer,
being the one at the end of inflation, even though it is entirely
evaluated at horizon crossing.

This convergence relies on all but one of the fields becoming
dynamically unimportant before inflation ends, which may be true of
typical trajectories but cannot be absolutely generic. In principle
one should then solve the problem numerically, but unfortunately this is
not tractable for the large number of fields that we are considering
(as one has to solve a trajectory for a separate perturbation in each
field direction\footnote{Recently Yokoyama et al.~\cite{YTSS}
developed a Wronskian-based approach which may improve the efficiency
of such multi-field calculations by identifying the perturbation
component contributing to the curvature perturbations, but in Nflation
there is no guarantee of a condition they require on convergence of
the trajectories.}), and so we adopt the slow-roll formula.

There is also the question of whether further perturbations might be
generated after inflation, for instance by a curvaton-like mechanism
(see e.g.~the discussion in Ref.~\cite{VW}). Such effects would be
absent if the late stages of inflation are driven by a single field,
but otherwise would depend on the routes by which the scalar fields
decay into conventional and dark matter. We assume such effects are
absent.

\subsection{Nflation perturbations: amplitude}

\begin{figure}[t]
\centering
\includegraphics[width=7.5cm]{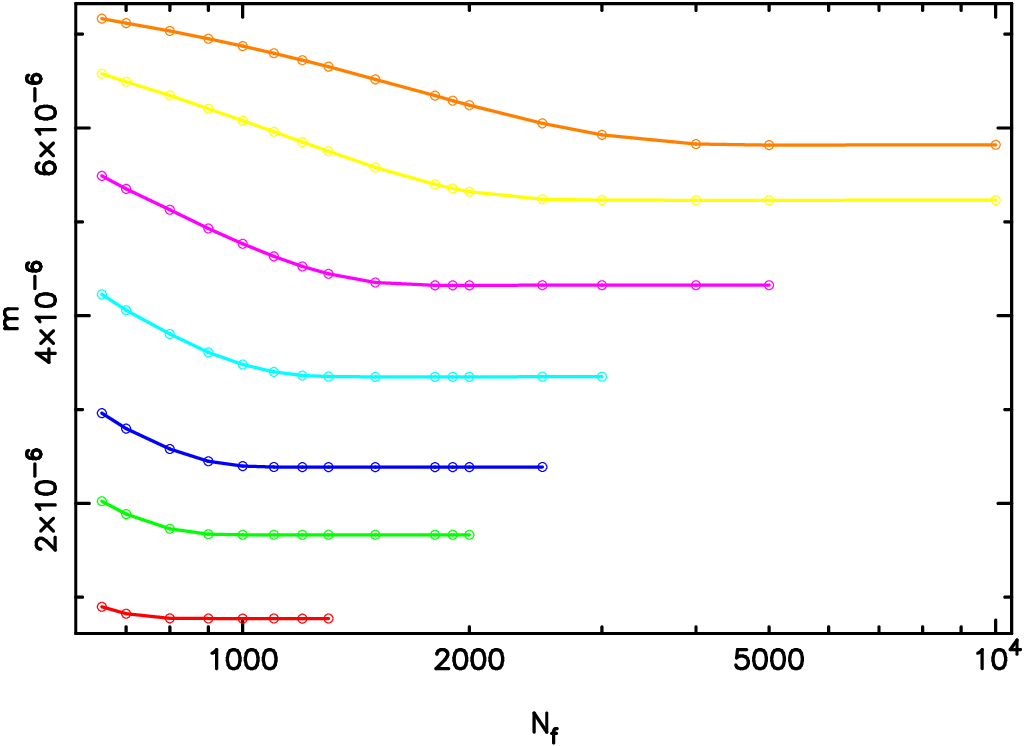}
\caption{\label{f:m} The mass $m$ of the lightest field, determined
from the normalization of the power spectrum.  This is shown as a
function of the number of fields $N_{{\rm f}}$ for a range of values
of $\sigma$ (from bottom to top, $\sigma = 100$, $150$, $200$, $300$, 
$500$, $1000$ and $2000$).}
\end{figure}

We first investigate the normalization of the mass spectrum enforced
by the perturbation amplitude. This is shown in Fig.~\ref{f:m} for a
set of values of $\sigma$. The single-field normalization is attained
in the limit $N_{{\rm F}} \ll \sigma$, in which the field packing
becomes very close and the equal mass case is attained, though only
the highest line shown here comes close to that case.

For a given value of $\sigma$, each curve flattens once $N_{{\rm f}}
\gg \sigma$. This indicates that the most massive fields are no longer
playing a role during the last 50 $e$-foldings of inflation, having
fallen to their minima before observable scales leave the
horizon. This flattening is a generic property of all observables.

\subsection{Nflation perturbations: spectral index}

\begin{figure}[t]
\centering
\includegraphics[width=7.5cm]{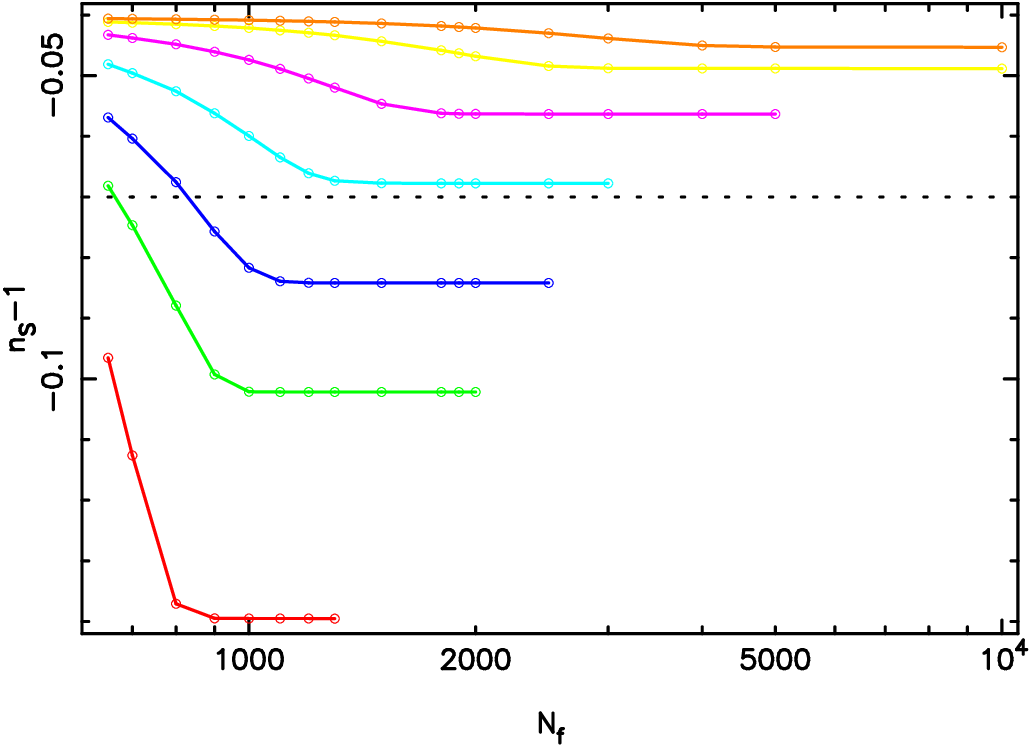}
\caption{\label{f:average} The predicted average value of the spectral
index as a function of the number of fields $N_{{\rm f}}$, shown for a
range of values of $\sigma$ (from bottom to top, $\sigma = 100$,
$150$, $200$, $300$, $500$, $1000$ and $2000$). The left-hand edge of
the graph roughly corresponds to the minimum number of fields needed
to achieve sufficient inflation, and the dotted line shows the
observational lower limit on $n_{{\rm S}}$ from WMAP3.}
\end{figure}

The most important observable is the spectral index, which we
calculate as described above and show in Fig.~\ref{f:average}. Each
point is the average value obtained over ten simulations with
different random initial conditions.  As with all observables, the
single-field value is attained in the limit $N_{{\rm f}} \ll \sigma$,
in which we reproduce the result of Byrnes and Wands \cite{BW} that
the perturbations match the single-field form.  But otherwise
significant differences are seen, which in some cases is enough to put
the model outside the region permitted by the WMAP3 data \cite{wmap3}.

Again we see the flattening of the curves for sufficiently large
$N_{{\rm f}}$, indicating that the most massive fields play no
dynamical role in the last 50 $e$-foldings.

The results shown in Fig.~\ref{f:average} are the mean values over
realizations of the initial conditions. However we also find that the
spread in $n_{{\rm S}}$ values is quite small; the standard deviation
of $n_{{\rm S}}$ is never more than a few percent of its displacement
from unity for all the values shown in that graph. The scatter in the
mass normalization shown in Fig.~\ref{f:m} is also at the few percent
level at most.  We conclude that with such large numbers of fields,
the observational predictions are to a good approximation {\em
independent} of the field initial conditions. The mass spectrum is
sufficiently dense that the space of initial conditions is well
explored in each realization, and a well-defined prediction for
$n_{{\rm S}}$ emerges in a manner analogous to the emergence of
macroscopic thermodynamic quantities.

We are now in a position to discuss what is permitted by observations,
using WMAP3 constraints in the inflationary $n_{{\rm S}}$--$r$ plane
\cite{wmap3}. We use limits obtained from WMAP3 data
alone.\footnote{The constraint plot in the existing (v1) WMAP3 paper
is known to be incorrect; we instead use limits quoted by Hiranya
Peiris in subsequent talks. See also Ref.~\cite{KKMR}.} All models
predict $r \simeq 0.16$, for which value the 95\% range on $n_{{\rm
S}}$ is $0.93 \leq n_{{\rm S}} \leq $1.02, only the lower limit
interesting us.

We see in Fig.~\ref{f:average} that this is a significant constraint
which is failed in large parts of parameter space. Provided $\sigma$
exceeds about 280, the spectral index is always large enough
regardless of the number of fields, provided that the lower limit
$N_{{\rm f}} \gtrsim 600$ to obtain sufficient inflation is
exceeded. As $\sigma$ reduces, an upper limit is then placed on the
number of fields. This limit rapidly comes into contradiction with the
sufficient $e$-foldings condition, so that by $\sigma =150$ there are
almost no viable models at all (this result presumes that no field value
exceeds the Planck mass). We conclude that Nflation is observationally
viable only if the fields are both numerous and very densely packed.

\section{Conclusions}

We have carried out a detailed investigation of the inflationary
dynamics and perturbations in Nflation models. The tensor-to-scalar
ratio is completely independent of the model parameters, i.e.~the
number of fields and their mass spectrum, and also independent of the
field initial conditions. Nflation therefore makes the definitive
prediction that $r=8/N$ \cite{AL}, where $N$ is the number of
$e$-foldings corresponding to when the comoving wavenumber at our
present Hubble radius equalled the Hubble radius during inflation
(typically $N \simeq 50$). That there is a unique prediction is a
feature only of the case of uncoupled fields with polynomial
potentials \cite{piao}. This prediction is readily testable by
upcoming experiments.

The spectral index, by contrast, is dependent on both the number of
fields, $N_{{\rm f}}$, and the density of fields in the mass spectrum,
$\sigma$. Provided $\sigma$ is large enough, it is however independent
of the initial conditions as they are statistically explored
sufficiently well by the densely-packed fields. This is analogous to
the emergence of macroscopic quantities in thermodynamics. For any
given $\sigma$, results also become independent of the number of
fields once it is large enough, as the most massive ones fall into
their minima before observable scales cross outside the horizon.

In terms of observations, if the density of fields is large enough,
$\sigma \gtrsim 280$, then a satisfactory spectral index is always
achieved regardless of the number of fields, provided this number
exceeds the minimum of around 600 needed to give sufficient
inflation. For smaller $\sigma$, compatibility with observations
imposes an upper limit on the number of fields, which rapidly becomes
incompatible with the condition for sufficient inflation, giving a
minimum $\sigma$ of 150 to have a successful scenario. Observations
therefore require the Nflation model both to have a large number of
fields and for these fields to have a very densely packed mass
spectrum.

\begin{acknowledgments}
S.A.K.\ was supported by the Korean government and A.R.L.\ by PPARC
(UK). We thank Richard Easther, David Lyth, Filippo Vernizzi, and
David Wands for useful discussions.
\end{acknowledgments}


\begin{thebibliography}{50}
\bibitem{DKMW} S. Dimopoulos, S. Kachru, J. McGreevy, and J. Wacker,
	{\tt hep-th/0507205}.
\bibitem{LMS} A. R. Liddle, A. Mazumdar, and F. E. Schunck, Phys.
	Rev. D{\bf 58}, 061301(R) (1998), {\tt astro-ph/9804177}.
\bibitem{KO} P. Kanti and K. A. Olive, Phys. Rev. D{\bf 60}, 043502
	(1999), {\tt hep-ph/9903524}; P. Kanti and K. A. Olive, 
	Phys. Lett. B{\bf 464}, 192 (1999), {\tt hep-ph/9906331}. 
\bibitem{KL} N. Kaloper and A. R. Liddle, Phys. Rev. D{\bf 61},
	123513, (2000), {\tt hep-ph/9910499}.
\bibitem{mfield} A. Jokinen and A. Mazumdar, Phys. Lett. B{\bf 597},
	222 (2004), {\tt hep-th/0406074}; K. Becker, M. Becker, and
	A. Krause, Nucl. Phys. B{\bf 715}, 349 (2005), {\tt 
	hep-th/0501130}.
\bibitem{EM} R. Easther and L. McAllister, JCAP {\bf 0605}, 018
	(2006), {\tt hep-th/0512102}.
\bibitem{AL} L. Alabidi and D. H. Lyth, JCAP {\bf 0605}, 016 (2006),
	{\tt astro-ph/0510441}
\bibitem{BW} C. T. Byrnes and D. Wands, Phys. Rev. D{\bf 73}, 063509 
	(2006), {\tt astro-ph/0512195}.
\bibitem{SS} M. Sasaki and E. D. Stewart, Prog. Theor. Phys. {\bf 95},
	71 (1996), {\tt astro-ph/9507001}.
\bibitem{LL} A. R. Liddle and D. H. Lyth, {\em Cosmological 
	inflation and large-scale structure}, Cambridge University Press, 
	Cambridge (2000).
\bibitem{LR} D. H. Lyth and A. Riotto, Phys. Rep. {\bf 314}, 1 (1999),
	{\tt hep-ph/9807278}.
\bibitem{VW} F. Vernizzi and D. Wands, JCAP {\bf 0605}, 019 (2006), 
	{\tt astro-ph/0603799}.
\bibitem{YTSS} S. Yokoyama, T. Tanaka, M. Sasaki, and E. D. Stewart, 
	{\tt astro-ph/0605021}.
\bibitem{wmap3} D. N. Spergel et al. (WMAP Collaboration), {\tt 
	astro-ph/0603449}; G. Hinshaw et al. (WMAP Collaboration), {\tt 
	astro-ph/0603451}; L. Page et al. (WMAP Collaboration), {\tt 
	astro-ph/0603450}.
\bibitem{KKMR} W. Kinney, E. W. Kolb, A. Melchiorri, and A. Riotto, 
	{\tt astro-ph/0605338}.
\bibitem{piao} Y.-S. Piao, {\tt gr-qc/0606034}.
\end{thebibliography}
\end{document}